\documentclass[acmjacm]{acmtrans2m}
\usepackage{times}
\usepackage{mathrsfs}
\usepackage{amsfonts}
\usepackage{amssymb}
\usepackage{amsmath}
\usepackage{amsmath,amssymb,color}
\usepackage{graphicx}
\usepackage{epsfig}
\usepackage{alg}
\makeatletter
\renewcommand\@biblabel[1]{\indent$[{#1}]$\quad\quad}
\makeatother

\newtheorem{theorem}{theorem}[section]

\newtheorem{lemma}[theorem]{Lemma}
\newdef{definition}[theorem]{Definition}
\newdef{remark}[theorem]{Remark}

\newcommand{\BibTeX}{{\rm B\kern-.05em{\sc i\kern-.025em b}\kern-.08em
    T\kern-.1667em\lower.7ex\hbox{E}\kern-.125emX}}

\title{Global Clock, Physical Time Order and Pending Period Analysis in Multiprocessor
Systems}
\author{\small{Yunji CHEN}\\ \footnotesize{\emph{Institute of Computing Technology, Chinese Academy of Sciences}}\\
\\\small{Tianshi CHEN}\\ \footnotesize{\emph{School of Computer
Science and Technology, University of Science and Technology of
China}}\\ \\\small{Weiwu HU}\\ \footnotesize{\emph{Institute of
Computing Technology, Chinese Academy of Sciences}}}

\begin{abstract}
In multiprocessor systems, various problems were treated with
Lamport's logical clock and the resultant logical time orders
between operations. However, one often needs to face the high
complexities caused by the lack of logical time order information in
practice.

In this paper, the so-called \emph{physical time order} is proposed
based on the \emph{global clock} in multiprocessor systems.
Concretely, we first utilize the global clock to infuse the
\emph{pending period} to each operation in a multiprocessor system,
where the pending period is a time interval in which the operation
starts and ends. Afterwards, we define the physical time order for
any pair of operations with disjoint pending periods. The physical
time order is an underlying characteristic of any real execution in
multiprocessor systems due to that it is part of the truly-happened
orders obeying real physical time. Formally, the physical time order
is proven to be independent and consistent with traditional logical
time orders.

The above novel yet fundamental concepts enables new effective
approaches for analyzing multiprocessor systems, which are named
\emph{pending period analysis} as a whole. As a consequence of
pending period analysis, many important problems of multiprocessor
systems can be tackled effectively. As a significant application
example, complete memory consistency verification, which was known
as an NP-hard problem, can now be solved with the complexity of
$O(n^2C^pp)$ by utilizing physical time order information (where $n$
and $p$ are the number of operations and processors respectively,
$C$ is some constant). Moreover, two event ordering problems, which
were proven to be Co-NP-Hard and NP-hard respectively, can both be
solved with the time complexity of $O(nC^pp)$ if restricted by
pending period information.
\end{abstract}





\begin{document}
\begin{bottomstuff}
%Corresponding author: Yunji Chen (Email: cyj@ict.ac.cn), Institute
%of Computing Technology, Chinese Academy of Sciences, P.O. Box
%2704-25, Beijing 100190, P. R. China.
Corresponding Address: Institute of Computing Technology, Chinese
Academy of Sciences, P.O. Box 2704-25, Beijing 100190, P. R. China.

Email: cyj@ict.ac.cn (Yunji Chen)

This paper is partially supported by the National High Technology
Development 863 Program of China under Grants No.2007AA01Z112 and
No.2008AA110901, the National Grand Fundamental Research 973 Program
of China under Grant No.2005CB321600, and the National Natural
Science Foundation of China under grant No. 60533020.
\end{bottomstuff}
\maketitle

\section{Introduction}\label{sec:introduction}
\subsection{Motivation and Related Work}\label{sec:motivation}
Many theoretical investigations of multiprocessor systems treat
processors as distributed spatially. In these investigations,
Lamport's logical clock \cite{Lamport1978}, which is known as a
cornerstone in the parallel and distributed computing areas, is
often utilized to partially order operations in multiprocessor
systems. Given a pair of operations obeying some logical time order
obtained by logical clock (such as processor order, execution order
and so on), the logically later operation should observe the effort
of the logically earlier operation. Once given the logical time
orders between all pairs of conflicting operations in a system, the
final execution result of the system has been determined. Thus the
logical order information, if perfect, can reveal some intrinsic
features of parallel and distributed computing without a global lock
which was considered hard-to-achieve. However, in real
multiprocessor systems, it is often the case that the logical order
information is far from perfect. That is because observing the
logical order information of all operations, especially the
low-level operations such as load, store, and synchronization
instructions, is often impractical in real systems. If purely
replying on fragmentary logical time order information, one have to
infer or conjecture the orders between a large number of conflicting
operations \cite{Netzer1990,Gibbons1994}. As a consequence, many
application problems in multiprocessor systems (e.g., memory
consistency verification \cite{Gibbons1992,Gibbons1994,Gibbons1997},
event ordering \cite{Netzer1990}, and so on) suffer from the
resultant high computational costs, and thus are hard to solve.

During the past decade of years, driving by the development of
integrated circuit process, SMP (Symmetric Multi Processors) and CMP
(Chip Multi Processor) techniques, the density of computing capacity
of multiprocessor systems is fast increasing. The resultant scaling
down of multiprocessor systems rewakes the intuitive idea of
utilizing global clock, and a number of investigations with the
consideration of global clock have been proposed. Herlihy and Wing
\cite{Herlihy1990} proposed the concept of linearizability, which
requires the accesses to the same memory location happening in
disjoint time intervals with respect to a global clock, as a
correctness condition of memory system. In \cite{Singla1997}, Singla
\emph{et al.} proposed a temporal memory model ``delta consistency''
to offer time window to coalesce write operations to the same memory
location. In \cite{Riegel2006,Spear2006}, the global counters, which
implicitly represent the global time, were employed to reason about
the ordering of transactions in transactional memory
\cite{Herlihy1993}. The common idea behind the above investigations
is to obtain logical order information (especially execution order
about the same memory location) by explicitly or implicitly
employing a global clock.

Nevertheless, the implication of a global clock is far more than
providing some complementary logical order information. A notable
fact is that, in a multiprocessor system, the global clock actually
provides a physical-time-based partial ordering containing all
happened operations in the system, and the ordering is ``nearly'' a
total order: Two operations are not ordered by this ordering if and
only if their precise performed times (the time when an operation is
observed by all processors) on the global clock are exactly the
same. The complementary logical order information obtained through
the global clock is only a part of the above partial ordering, since
the logical orders concern merely operations on the same processor
or accessing the same location. In fact, the extra order information
obtained through the global clock, together with other logical order
information, can produce a transition closure that further extends
the acquirable order information. To our best knowledge, few
investigation has concerned the above extra order information. Such
neglect is probably due to the traditional view that ``logical time
orders are enough to determine the result of an execution, thus
concerning extra order information , which does not change the
result of execution, is not necessary''. However, in this paper, it
is discovered for the first time that the extra order information is
powerful for simplifying many problems in multiprocessor systems.

Although we have mentioned the motivation of the paper above, to
make better use of global clock in multiprocessor systems, we will
still encounter three concrete difficulties in practice: First, it
is hard to obtain the precise performed time of an operation, even
there is a precise physical global clock. The reason is that the
exact performed time of an operation is correlated with the
hard-to-observe inner states of all processors and the network of
the multiprocessor system (e.g., whether an invalidation message has
arrived at some processor). Second, even if one makes some
compromise and utilizes a time interval which includes the precise
performed time, instead of the precise performed time itself, one
must face new problems, e.g., how to deal with the overlapped time
intervals of logically ordered operations (noting that most modern
processors can execute multi operations overlappedly)? Third, it may
be difficult to observe and record the global time information for
all operations, since in a modern multiprocessor system there are
too many operations performed in every second. In this paper, these
difficulties are tackled effectively and efficiently by our
approaches.

\subsection{Our Contributions}
Given a global clock of a multiprocessor system in which the effect
of an operation can always be globally observed in bounded
time\footnote{For the sake of brevity, when we are talking about a
multiprocessor system, we imply that the precondition does hold.},
let us consider two bounding time points for the performed time of
each operation, named the start time and the end time. The resultant
time interval from the start time to end time, which includes the
performed time, is called the \emph{pending period}. As a relaxation
of the performed time, the pending period is easier to obtain in
comparison with the precise performed time, which will be shown in
Section \ref{sec:Assignment Analysis} in detail. It is worth noting
that the concept of pending period is pervasive: The pending periods
of two operations in the same processor can be overlapped, which
enables the instruction-level parallelism (ILP) \cite{Jouppi1989}
adopted by most modern processors; the pending periods of two
operations accessing a same memory location can also be overlapped,
which leaves a space of overlapping memory accesses for efficiency
optimization. Thus the concept of pending period can be adopted in
most memory models, from strong memory consistency models (such as
linearizability \cite{Herlihy1990}, sequential consistency
\cite{Lamport1979,Scheurich1987}), to weak memory consistency models
(such as weak consistency \cite{Dubois1986}, release consistency
\cite{Keleher1992}).

On the other hand, once any two operations have disjoint pending
periods (even they relate to neither the same processor nor the same
memory location), there is a \emph{physical time order} between the
two operations. In Section \ref{sec:Physical Time Order}, it is
proven that physical time order is independent and consistent with
existing logical time orders. This is not surprising since the
physical time order is part of a truly-happened physical-time-based
ordering with respect to the global clock. Thus the physical time
order is a natural order which must be obeyed by any real execution
in multiprocessor systems.

On the basis of global clock, pending period and physical time
order, we introduce some effective approaches for tackling problems
in the context of pending period, which are named \emph{pending
period analysis} as a whole. These approaches attempt to utilize the
information brought by global clock to the full extent, involving
but not limited to the traditional logical time order information.
The first approach, assignment analysis, aims at assigning values
for pending periods of all operations when the pending periods of
only part of operations can be observed directly. This approach can
effectively handle the difficulty of observing and recording an
immoderate amount of operations by inferring pending periods for
part of operations. The second approach called frontier analysis
distinguishes the operations with overlapped pending periods from
those with disjoint pending periods, and manage to prune the
frontier graph \cite{Gibbons1994}. As a consequence, the maximal
number of nodes in a frontier graph is significantly reduced from
$O((n/p)^p)$ to $O(nC^p)$, while the maximal number of edges is
reduced from $O((n/p)^{p+1})$ to $O(nC^pp)$. Noting that many
problems in multiprocessor systems can come down to graphic problems
on the frontier graph, frontier analysis may be applicable to reduce
the time complexities of these problems. The third approach, order
analysis, further explores orders beyond the physical time order.
The order analysis aims at characterizing the so called \emph{time
global order}, which is the transition closure of the physical and
logical time orders. A result of order analysis is that any cycle,
in the \emph{TGO execution graph} representing the time global
orders in a system, can be localized to involve merely $O(p)$
operations. This conclusion is important to guarantee the
correctness of a multiprocessor system.

One established example for validating the effectiveness of physical
time order and pending period analysis is the well-known memory
consistency verification problem, which was known as an NP-hard
problem \cite{Gibbons1992,Gibbons1994}. With the concept and
approaches developed in this paper, the problem can be solved with
the time complexity of $O(n^2C^pp)$ in the context of pending
period. This method has been employed in validation of an industrial
CMP \cite{Chen2009,Hu2009}. Additional examples are the event
ordering problems which investigate the possible orders between
pairs of operations. The investigated problems have been proven to
be co-NP and NP respectively \cite{Netzer1990}. However, if these
problems are restricted by physical time order, then they can be
solved with the time complexity of $O(nC^pp)$. The successful
applications of our approach demonstrates that the global clock,
physical time order and pending period analysis are effective and
efficient in tackling various problems in multiprocessor systems,
especially those problems relating to frontier graph.

\begin{table*}[htbg]
\centering
\begin{tabular}{|ll|}
\hline
$\mathcal{S}$ & Multiprocessor system\\
$u,v$ (with a subscript) & Operation\\
$w$ (with a subscript) & Write operation\\
$r$ (with a subscript) & Read operation\\
$O$ (with a subscript) & Set of operations\\
$\mathcal{P}$ (with a subscript) & Process / processor\\
$P$ (with a subscript) & Path on a graph\\
$\mathbb{P}$ & Set of paths on a graph\\
$\mathscr{C}$ & Set of cycles on a graph\\
$\mathcal{C}$ & Cycle on a graph\\
$f$ & Frontier\\
$n$ & Number of operations\\
$p$ & Number of processes (processors)\\
$C$ & Constant\\
$E$ & Execution order\\
$P$ & Program order\\
$PO$ & Processor order\\
$GO$ & Global order\\
$T$ & Time order\\
$TGO$ & Time global order\\
\hline
\end{tabular}
\caption{Notations in this paper.\label{table:notations}}
\end{table*}

The contributions of this paper can be summarized as follows: First,
we provide a novel notion for global clock in multiprocessor
systems, showing that the physical-time-based partial ordering
information exported from the global clock, though has been
neglected in some sense, is very powerful in tackling many problems
in multiprocessor systems. Second, the proposed physical time order
has established a novel, natural and fundamental concept for
multiprocessor systems, which is independent but consistent with the
traditional logical time orders. Third, a set of approaches, called
pending period analysis as a whole, are developed and have been
successfully used in two types of well-known application problems in
multiprocessor systems, one of which has been employed in industry.
The resultant new solutions for the application problems have made
significant improvements in comparison with the previous results.

The terminology used in the rest of this paper is introduced in
Table \ref{table:notations}. The rest of the paper is organized as
follows: Section 2 provides the definition of physical time order in
multiprocessor systems. Section 3 presents the approaches belonging
to pending period analysis. Section 4 introduces two related
applications. Section 5 concludes the whole paper.

\section{Physical Time Order in Multiprocessor Systems}
Lamport's logical clock \cite{Lamport1978} enables partially
ordering the events (operations) occurring at different processes
(processors) of a distributed system. During the past thirty years,
his logical ``happened-before'' order has generated different types
of logical time orders for multiprocessor systems, such as processor
order, execution order and so on. Briefly, if we are able to know
all the ``happened-before'' orders (i.e., logical time orders)
between operations, the execution of these operations, which can be
represented by a DAG (Directed Acyclic Graph) named execution graph,
is then determined. However, in practice it is often the case that
we can only acquire part of the logical time orders between
operations. Hence, if purely replying on the logical clock, one may
need to infer and conjecture the orders between pairs of operations,
which may be inherent with enormous search spaces. As a result,
there might be an intractably large number of candidate executions
that do not violate our obtainable information, which may be not
only incorrect but also difficult to distinguish the incorrectness.

An intuitive idea of tackling the above situation is to rewake the
idea of global clock. Probably, the simplest way of utilizing global
clock is to obtain the precise performed time of all operations.
However, due to that observing the precise performed time of one
operation may involve many components in the system, this ideal
approach is rather impractical. An alternative choice is to relax
the precise performed time to the physical time interval (i.e., the
pending period) which includes the performed time. In this way,
although the compromise may eliminate some order information implied
by global clock, partially allocating and utilizing the natural
physical ``happened-before'' order implied by the physical time
become possible. Such an order is called physical time order.

In this section, we will first provide a brief introduction to
traditional logical time orders. After that, we will provide more
detailed explanations for pending period and physical time order,
including the corresponding definitions. Finally, the relationship
between the physical and logical time orders is studied
theoretically.

\subsection{Logical Time Orders in Multiprocessor Systems}
In this subsection, we briefly introduce some traditional logical
time orders in multiprocessor systems, which are based on logical
clocks. Concretely, let us review the definitions of program order,
processor order and execution order, which are three well-known
types of logical time orders in multiprocessor systems.

\begin{definition}[(Program Order)]
Given two different operations $u_1$ and $u_2$ in the same
processor, we say that $u_1$ is before $u_2$ in program order iff
$u_1$ is before $u_2$ in the program. We denote this as
$u_1\xrightarrow[]{P}u_2$.
\end{definition}

\begin{definition}[(Processor Order)]
Given two different operations $u_1$ and $u_2$ in the same
processor, we say that $u_1$ is before $u_2$ in processor order iff
there is global agreement that $u_1$ is before $u_2$ for all
processors. We denote this as $u_1\xrightarrow[]{PO}u_2$.
\end{definition}

In multiprocessor systems, two operations are conflicting if they
access the same memory location and at least one of them is a store
operation \cite{Shasha1988}. The execution order specifies the order
between two consequent conflicting operations \cite{Hu2001}.

\begin{definition}[(Execution Order)] We say that a write operation
$w$ is before operation $u$ in execution order iff $w$ is the latest
write operation before $u$ that accesses the same memory location as
$u$. We denote this as $w\xrightarrow[]{E}u$. We say that a write
operation $w$ is after operation $u$ in execution order iff $w$ is
the first write operation after $u$ that accesses the same memory
location as $u$. We denote this as $u\xrightarrow[]{E}w$.
\end{definition}

In addition, the transitive closure of processor and execution
orders is known as the global order:

\begin{definition}[(Global Order)] We say that operation
$u_1$ is before operation $u_2$ in global order iff $u_1$ is before
$u_2$ in processor order, or $u_1$ is before $u_2$ in execution
order, or $u_1$ is before some operation $u$ in global order and $u$
is before $u_2$ in global order. Formally,
\begin{eqnarray}
\nonumber &&(u_1\xrightarrow[]{GO}u_2) \rightarrow
\big((u_1\xrightarrow[]{PO}u_2)
    \vee (u_1\xrightarrow[]{E}u_2) \vee(\exists u\in O:
u_1\xrightarrow[]{GO}u\xrightarrow[]{GO}u_2) \big).
\end{eqnarray}
\end{definition}

So far we have already introduced a number of traditional logical
time orders in multiprocessor systems. All the above orders are
based on Lamport's ``happened-before'' logical relation
\cite{Lamport1978}: the logically former operation is before the
logically latter operation in logical time order. In practice, it is
often the case that only parts of those relations can be observed
directly, especially the execution orders \cite{Hangal2004}. For
example, to observe the write-write execution orders, one may add
specific hardware to cache coherence maintainer \cite{Meixner2005}.
Even if we can infer some hard-to-observe orders based on known
orders, the number of candidate executions for parallel programs may
still be intractably large, since the case in which we can infer all
logical time orders is rare. To cope with this problem by exploiting
more information about the relations between operations, we propose
the so-called physical time order in the next subsection.

\subsection{Physical Time Order}\label{sec:Physical Time Order}
In a von Neumann architecture, an operation must be fetched into the
processor before it is executed, hence there is a start time for any
operation. The above fact implies that, before an operation starts,
it cannot affect other operations. On the other hand, an operation
can be globally observed before it ends at some bounded time point,
given some appropriate hardware (e.g., store atomicity
\cite{Arvind2006}) or supplementary software (e.g., broadcasting).
Therefore, given the global clock, we can assign a start time and an
end time to each operation in a multiprocessor system. Compared with
the performed time which involves the states of all processors, all
caches and the network in the whole multiprocessor system, the start
time and the end time are easier to obtain because of their locality
and flexibility.

Based on the start time and the end time of an operation, we provide
the following definition of \emph{pending period}.

\begin{definition}[(Pending Period)]
The pending period of $u$ is the period from $t_s(u)$ to $t_e(u)$.
We say that an operation $v$ is in the pending period of operation
$u$ iff the pending periods of the two operations are overlapped.
\end{definition}

The concrete methods of observing start and end times are
architecture-dependent. Hence, there are different observations and
implementations for pending period and the related supports.
However, the following essential idea of pending period is
invariable: \emph{The performed time of an operation --i.e., the
time when the operation is performed globally-- must be in its
pending period.} Therefore, regardless of the concrete definitions
of start time and end time, a partial order exists between two
operations executing in disjoint pending periods. We call the
partial order \emph{physical time order}.

\begin{definition}[(Physical Time Order)]\label{physical_time_order}
Given two operations $u$ and $v$, if the following
\begin{enumerate}
\item the performed times of $u$ and $v$ are in their pending
periods respectively;
\item the end time of operation $u$ is before the start time of operation
$v$
\end{enumerate}
both hold, then we say that $u$ is before $v$ in physical time
order. Formally,
\begin{eqnarray}
\big(t_s(u)\leq t_p(u)\leq t_e(u)\big)\wedge \big(t_s(v)\leq
t_p(v)\leq t_e(v)\big)\wedge \big(t_e(u)< t_s(v)\big)\leftrightarrow
(u\xrightarrow[]{T}v).
\end{eqnarray}
\end{definition}

\begin{figure}[htbp!]
\label{figure:pending period}
\begin{center}
\epsfig{file=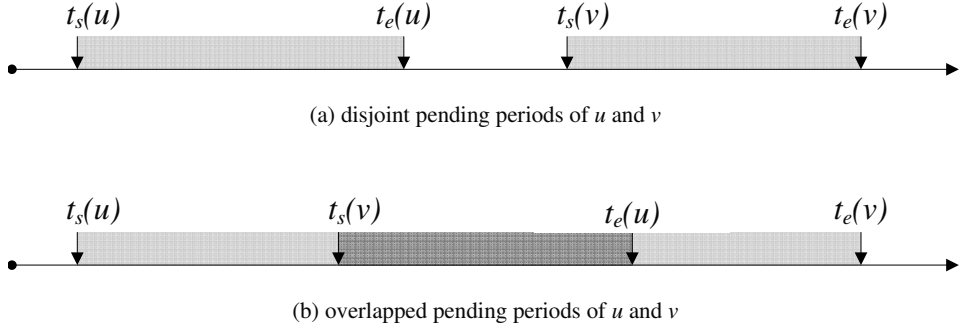, width=1\textwidth}
\end{center}
\caption{Time points of operation $u$ and $v$ on time axis (left
side is earlier), where $t_s(\cdot)$ and $t_e(\cdot)$ represent
start time and end time respectively.}
\end{figure}

As illustrated in Figure \ref{figure:pending period}a, the pending
periods of $u$ and $v$ are disjoint, thus there is physical time
order between $u$ and $v$, i.e., $u\xrightarrow[]{T}v$. In Figure
\ref{figure:pending period}b, the pending periods of $u$ and $v$ are
overlapped. Hence, either the performed time of $v$ is before the
performed time $u$, or the performed time of $u$ is before the
performed time $v$ are possible. Therefore, in such a case there is
no physical time order between $u$ and $v$. That is, if operation
$v$ is in the pending period of operation $u$, then
$\neg(u\xrightarrow[]{T}v \vee v\xrightarrow[]{T}u)$ holds according
to the above definition.

Notably, the physical time order between $u$ and $v$ does not
require that the two operations are executing in the same processor
or accessing the same memory location. Instead, it simply depends on
the pending periods of $u$ and $v$ resulted from the physical time
given by the global clock.

\subsection{Relationship between Physical and Logical Time Orders}\label{sec:Relationship between Physical and Logical Time Orders}
In this subsection, we discuss the relationship between physical
time order and traditional logical time orders. Fundamentally, a
remarkable difference between the former and the latter is that,
physical time order is based on physical global clock while logical
time orders are based on logical clock. Due to the above difference,
it is intuitive that physical time order is independent with logical
time order:

\begin{theorem}[Time Order Independency Theorem]\label{Th:time_order_independency}
The physical time order and logical time order are independent with
each other.
\end{theorem}
\textbf{\emph{Proof.}} First, let us consider a multiprocessor
system in which the start times and the end times of all operations
are $0$ and $\infty$ respectively. In such a system, there is no
physical time order between any pair of operations. However, there
can be some logical time order between operations. Therefore, the
physical time order does not contain any logical time order.

Second, let us consider a multiprocessor system in which all
operations have nontrivial assignments of pending periods, and
access distinct memory locations. In such a system, there is no
logical time order between any two operations in different
processors, while there can be some physical time order between
operations in different processors. Therefore, any logical time
order does not contain the physical time order. $\square$

Another issue concerns the following question: Whether the physical
time order and logical time order contradict with each other? As we
know, the physical time order between two operations implies that
one operation \emph{physically} happens before the other operation,
while the logical time order between two operations implies that one
operation \emph{logically} happens before the other operation.
Intuitively, if there is no bug in the multiprocessor system, the
logical clock should comply with the physical global clock.

To carry out the investigation, we need a common and natural
criteria for comparing physical time order with logical time orders.
A solution is to use the physical time points given by global clock
to characterize all the above orders, since physical time order is
based on the start and end time points, and logical time order can
be represented as the relation between the performed time points.
Following this idea, we list all formulae linking the orders to
physical time points as follows:

\begin{enumerate}
\item The performed time of operation $u$ is between the start time and the end time
of $u$:
\begin{eqnarray}\nonumber t_s(u)\leq t_p(u)\leq t_e(u);
\end{eqnarray}
\item If $u$ is before $v$ in physical time order, then the end time of $u$ is
before the start time of $v$: \begin{eqnarray}\nonumber
(u\xrightarrow[]{T}v)\rightarrow \big(t_e(u)< t_s(v)\big);
\end{eqnarray}
\item If $u$ is before $v$ in processor order, then the performed time of $u$ is
before the performed time of $v$: \begin{eqnarray}\nonumber
(u\xrightarrow[]{PO}v)\rightarrow \big(t_p(u)< t_p(v)\big);
\end{eqnarray}
\item If $u$ is before $v$ in execution order, then the performed time of $u$ is
before the performed time of $v$: \begin{eqnarray}\nonumber
(u\xrightarrow[]{E}v)\rightarrow \big(t_p(u)< t_p(v)\big);
\end{eqnarray}
\end{enumerate}

Based on the above formulae, we can prove the following theorem.

\begin{theorem}[Time Order Consistency Theorem] The physical time
order is consistent with the global order, i.e., the transition
cloture of the processor and execution orders. Formally,
\begin{eqnarray}
(v\xrightarrow[]{T}u)\rightarrow \neg (u\xrightarrow[]{GO}v).
\end{eqnarray}
\end{theorem}
\textbf{\emph{Proof.}} According to Definition
\ref{physical_time_order}, the above theorem is equivalent to
$(u\xrightarrow[]{GO}v)\rightarrow \big(t_s(u)< t_e(v)\big)$. Since
$(u\xrightarrow[]{PO}v)\rightarrow \big(t_p(u)< t_p(v)\big)$ and
$(u\xrightarrow[]{E}v)\rightarrow \big(t_p(u)< t_p(v)\big)$ both
hold, by transitivity of partial order we obtain that
$(u\xrightarrow[]{GO}v)\rightarrow \big(t_p(u)< t_p(v)\big)$. Since
$t_s(u)\leq t_p(u)$ and $t_p(u)\leq t_e(u)$,
$(u\xrightarrow[]{GO}v)\rightarrow \big(t_s(u)< t_e(v)\big)$ holds.
$\square$

The independency and consistency between the physical time order and
traditional logical time orders demonstrate that the former is novel
yet natural. In the rest part of the paper, we will show that
physical time order, together with our forthcoming approaches named
pending period analysis, are powerful for tackling many problems in
multiprocessor systems.

\section{Pending Period Analysis}
So far we have introduced the concepts of pending period and
physical time order in the context of global clock. In this section,
three concrete approaches based on the aforementioned concepts are
proposed to analyze the problems in multiprocessor systems. These
approaches are named pending period analysis as a whole, since they
concentrate on differen aspects of pending periods.

The first approach \emph{assignment analysis} aims at obtaining the
pending periods of operations. As we have mentioned, the physical
time order does exist between two operations only if the pending
periods of the two operations are disjoint, thus knowing the pending
periods of operations is essential to obtaining the potential
physical time order between operations. Moreover, as a new
restriction to multiprocessor systems, pending period may forbid
many candidate executions of parallel programs which violate the
ordering imposed by a global clock. Accordingly, the second
approach, named \emph{frontier analysis}, aims at pruning the search
space of candidate executions in the context of physical time orders
and pending periods. Finally, suppose we have already known the
pending periods of all operations by some approach, then there are
probably some operations between which no physical time order holds
in response to their overlapped pending periods. The third approach,
\emph{order analysis}, aims at tackling the undiscovered order for
the operations with overlapped pending periods. In this section, the
above approaches will be introduced in detail.

\subsection{Assignment Analysis: Inferring Pending Periods}\label{sec:Assignment Analysis}
To obtain all the physical time orders between operations, one must
know the pending periods (determined by the start and end times) of
all operations in a multiprocessor system. As we have mentioned in
Section \ref{sec:Physical Time Order}, there are different ways to
observe pending period for facilitation in different systems. For
example, some dedicated registers are added to each processor in
Godson-3 \cite{Chen2009} to observe the bounding time points of
instructions (low-level operations), including
program\_counter/start\_time pairs of the last started operation and
program\_counter/end\_time pairs of the last ended operation
\cite{Chen2009}. Purely software method is also possible to observe
time points especially for high-level operations, e.g., employing
some global memory address as time counter.

On the other hand, in many cases, it is hard to observe directly the
pending periods of all operations if there are too many operations
(especially low-level operations such as load, store, and
synchronization instructions). Furthermore, even one can observe all
the pending period information, there may also be difficult to
record all pending period information in many systems if the speed
of generating pending period information is faster than the speed of
recording pending period information.

To cope the difficulty of obtaining and recording pending period
information, the assignment analysis is proposed to make a
compromise by observing the pending periods for only part of
operations, and inferring the pending periods for the rest
operations according to the following rule:

\textbf{Inferred Pending Period Rule}\footnote{In some
multiprocessor systems supporting some weaker consistency other than
sequential consistency, there may be no total ordering for the
operations on a single processor. Therefore, the $\emph{prec}_{obs}$
or $\emph{succ}_{obs}$ of $u$ may be not unique. However, arbitrary
$\emph{prec}_{obs}$ and $\emph{succ}_{obs}$ for $u$ are reasonable
to provide a legal assignment of pending period to $u$.}: In a
multiprocessor system, given an operation set
$O=\{u_1,u_2,\dots,u_n\}$, and an observed operation subset
$O_{obs}=\{u_{obs_1},u_{obs_2},\dots,u_{obs_m}\}$ ($1\le
obs_1<obs_2<\dots<obs_m\le n$), if the pending period
$[t_s(u_{obs_j}),t_e(u_{obs_j})]$ has been observed for each
operation $u_{obs_j}\in O_{obs}$, then the pending period for any
operation $u\in O \setminus O_{obs}$, denoted by $[t_s(u),t_e(u)]$,
can be inferred as follows:
\begin{eqnarray*}
&t_s(u)=t_s(\emph{prec}_{obs}(u))\\
&t_e(u)=t_e(\emph{succ}_{obs}(u))
\end{eqnarray*}
where $\emph{prec}_{obs}(u)$ and $\emph{succ}_{obs}(u)$ are the last
observed operation before $u$ in processor order and the first
observed operation after $u$ in processor order
respectively\footnote{To avoid there is no $\emph{prec}_{obs}$ or
$\emph{succ}_{obs}$ for any unobserved operation, we require that
each operation without predecessor or successor operations in
\emph{processor order} should be observed.}. Formally,
\begin{eqnarray*}
&(\emph{prec}_{obs}(u)\in O_{obs})\wedge
(\emph{prec}_{obs}(u)\xrightarrow[]{PO}u) \wedge \nexists
j(\emph{prec}_{obs}(u)\xrightarrow[]{PO}u_{obs_j}\xrightarrow[]{PO}u)\\
&(\emph{succ}_{obs}(u)\in O_{obs})\wedge
(u\xrightarrow[]{PO}\emph{succ}_{obs}(u)) \wedge \nexists
j(u\xrightarrow[]{PO}u_{obs_j}\xrightarrow[]{PO}\emph{succ}_{obs}(u))
\end{eqnarray*}

We call a pending period of an operation is legal if it contains the
performed time of the operations. It is not difficult to prove the
legality brought by inferred pending period rule.

\begin{theorem}[Inferred Pending Period Theorem] If the observed pending period of
each observed operation is legal, then the inferred pending period
based on the inferred pending period rule is legal.
\end{theorem}
\textbf{\emph{Proof.}} Consider an operation $u\in O \setminus
O_{obs}$. If the inferred start time of $u$ is $0$, then
$t_s(u)=0\leq t_p(u)$, else $t_s(u)=t_s(\emph{prec}_{obs}(u))\leq
t_p(\emph{prec}_{obs}(u))< t_p(u)$. Hence, $t_s(u)\leq t_p(u)$
holds. If the inferred end time of $u$ is $\infty$, then $t_p(u)<
\infty t_e(u)$, else $t_p(u)< t_p(\emph{succ}_{obs}(u))\leq
t_e(\emph{prec}_{obs}(u))= t_e(u)$. Hence, $t_p(u)\leq t_e(u)$
holds. Since the performed time of $u$ is in the inferred pending
period of $u$, the inferred pending period is legal. $\square$

\begin{figure}[htbp!]
\label{figure:inferred time points}
\begin{center}
\epsfig{file=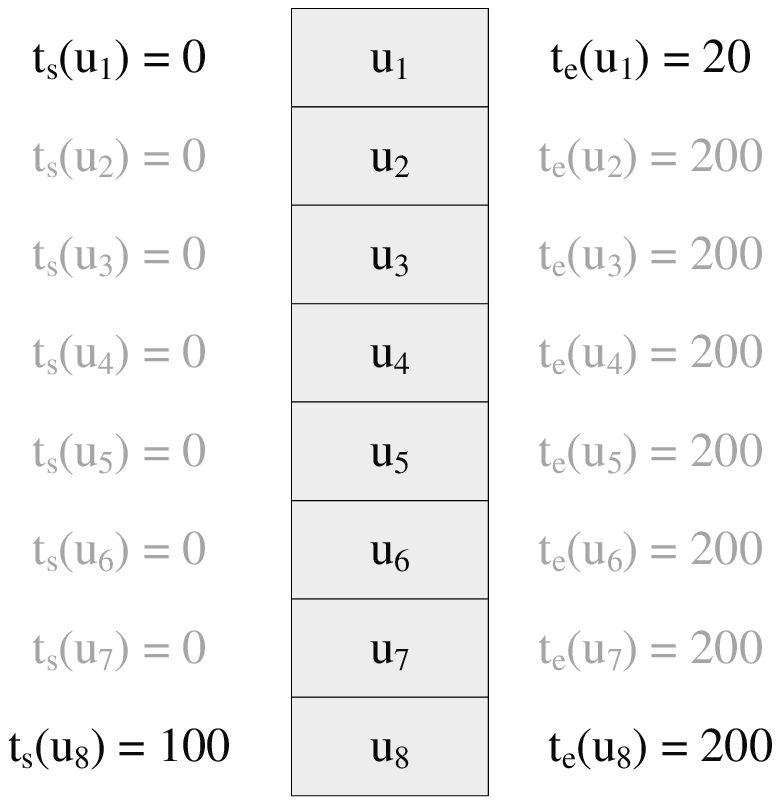, width=0.5\textwidth}
\end{center}
\caption{Time points of operations $u_2$ to $u_7$ are inferred with
$u_1$ and $u_8$.}
\end{figure}

As shown in Figure \ref{figure:pending period}, with merely observed
pending periods of $u_1$ and $u_8$, the pending periods of $u_2$ to
$u_7$ can be inferred: their start times are the same with the start
time of $u_1$, their end times are the same with the end time of
$u_8$.

Considering $n$ operations, with assignment analysis, one only need
to observe directly the pending period of one operation out of every
$m$ operations (for example, one out of every 100 (i.e., $m=100$)
operations is observed directly in \cite{Chen2009}), the pending
periods of the rest $n(m-1)/m$ operations can be then inferred. As a
result, the inferred pending period would be looser than the
observed pending period. However, this approach can reduce the
difficulty of observing and recording pending period information by
inferring pending periods.

\subsection{Frontier Analysis: Pruning Frontier Graph}\label{sec:Frontier Analysis}
For multiprocessor systems, it is a natural problem to analyze the
candidate executions of a parallel program. When the logical time
order information is ``perfect'', i.e., the orders between any pairs
of operation (if there are), are known to us, the above problem is
trivial. However, since it is often the case that we can only
observe and infer limited logical time order information (especially
execution order), we may need to consider a number of candidate
executions with respect to the same program, which do not conflict
the obtainable order information. If the available order information
is not enough, some execution may be actually \emph{illegal}
(violating memory consistency or cache coherence) but it has been
considered as candidate executions. Unfortunately, distinguishing
such an illegal execution from the legal one might be very
time-consuming \cite{Gibbons1992,Hangal2004}. In this section, we
propose an effective and efficient approach, which is named frontier
analysis, to prune the space of candidate execution for a parallel
program. The frontier analysis approach is based on the
aforementioned concept of physical time order and pending period. To
present our approach, we begin with Gibbons and Korach's notion of
frontier graph \cite{Gibbons1994}. Following the brief introduction
to frontier graph, we then show how frontier graph is pruned in
terms of frontier analysis. A corresponding complexity description
will also be provided.

Briefly, let us introduce Gibbons and Korach's concept of frontier
graph: Given $p$ processes $\mathcal{P}_1,\dots,\mathcal{P}_p$, and
$p$ sets $O_1,\dots,O_p$ containing all the operations executing at
the above $p$ processes respectively. A frontier $f(u_1,\dots,u_p)$
is a tuple of $p$ operations, where $\forall i\in \{1,\dots,p\}$,
$u_i\in O_i$. Gibbons and Korach \cite{Gibbons1994} further proposed
the frontier graph consisting of all possible frontiers in the
system and the directed edges connecting frontiers. It starts with
an \emph{starting frontier} consisting of $p$ NULL operations, which
represents the situation that no operation has executed at the very
beginning. It ends with a \emph{terminating frontier} consisting of
the terminating operations at the $p$ processes, which represents
the situation that all operations has begun at the very end. If a
new operation $u_i'\in O_i$ happens, the frontier $f(u_1,\dots,
u_i,\dots,u_p)$ will be updated to another frontier $f'(u_1,\dots,
u_i',\dots,u_p)$, and from $f$ to $f'$ there is a directed edge in
the frontier graph. Therefore, each candidate execution can be
mapped to a path from the starting frontier to the terminating
frontier on the frontier graph.

In multiprocessor systems, a frontier can demonstrate a snapshot of
the executing operations in the $p$ processors: the executing
operations in the processors $\mathcal{P}_1,\dots, \mathcal{P}_p$
are $u_1,u_2,\dots,u_p$ respectively. Intuitively, many important
problems of multiprocessor systems, which is inherent with search
spaces of candidate executions, can be transformed to graphic
problems related to the frontier graph, such as memory consistency
verification and event ordering problems. The complexities of
solving these problems directly relate to the size of frontier
graph. Unfortunately, according to Gibbons and Korach
\cite{Gibbons1994}, there are $O((n/p)^p)$ possible frontiers in
total, which results in the intractability of many multiprocessor
system problems. In \cite{Gibbons1994}, Gibbons and Korach proposed
to use additional information such as read mapping (the mapping from
every read to the write sourcing its value) and total write order
(the write order for each memory location totally) to simplify the
traverse on the frontier graph. Intuitively speaking, read mapping
and total write order can reduce the number of possible edges
connecting to each frontier by specifying the relations between
write and read operations, and it can also reduce the number of
reachable frontiers. Consequently, the time for finding a path from
the starting frontier to the terminating frontier on the frontier
graph is also reduced. However, since read mapping and total write
order may involve all processors in a system, observing and
restricting read mapping and total write order (especially the
latter) might be difficult in practice \cite{Hangal2004}.

\begin{figure}[htbp!]
\label{figure:frontier}
\begin{center}
\epsfig{file=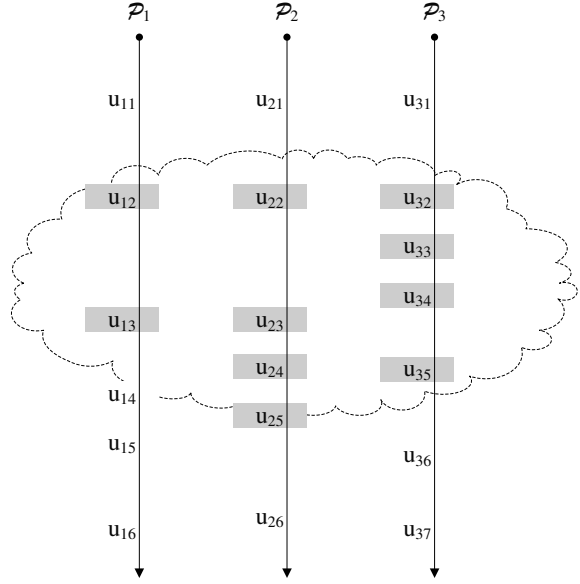, width=0.6\textwidth}
\end{center}
\caption{The operations in the frontier involving operation $u_{23}$
should be in the pending period of $u_{23}$.}
\end{figure}

Instead of relying on some logical information about read mapping
and total write order, we infuse the natural information including
physical time orders and pending periods into the frontier graph,
and manage to prune the frontier graph in frontier analysis. The
idea of frontier analysis is quite straightforward: Since the
operations in the same frontier are executing overlappedly, an
operation in a frontier is in the pending periods of other
operations belonging to the same frontier. Meanwhile, the physical
time orders eliminate the frontiers which contain two or more
operations with disjoint pending periods. As the example shown in
Figure \ref{figure:frontier}, the cloud represents the pending
period of operation $u_{23}$. Only the shadowed operations in the
clouds can appear in the frontiers involving $u_{23}$, since the
operations out of the pending period of $u_{23}$ cannot execute
overlappedly with $u_{23}$.

With the above natural property, we can again figure out the number
of feasible frontiers (in the context of physical time order) which
contain a specific operation $u_i$ in a $p$-processor system. A
feasible frontier containing $u_i$ can be obtained by picking $p-1$
operations from the operations whose pending period is overlapped
with that of $u_i$, to fill them to the corresponding seats of the
frontier. Moreover, since each operation in the systems discussed in
this paper can be globally observed in bounded time, we let $B$ be
the upper bound of the lengths of all pending periods. Hence, in one
processor, the maximal number of operations within any pending
periods is no larger than $cB$ (where $c$ is a constant, and we let
$C=cB$), which dose not rely on the number of operations $n$.
Therefore, there can be at most $C^{p-1}$ feasible frontiers
involving $u_i$. Meanwhile, since we have at most $n$ different
choices when determining the specific operation $u_i$, the total
number of feasible frontiers in the frontier graph is then
significantly reduced from $O((n/p)^p\big)$ to $O(nC^p)$. Noting
that the number of processors, $p$, is a constant for a given
multiprocessor system, thus the number of feasible frontiers is
actually $O(n)$ for a given system.

Further, the number of edges in the pruned frontier graph can also
be calculated. As we know, there are $p$ operations in a frontier,
and we know that within the pending period of a specific operation
there are at most $Cp$ operations in the total $p$ processors.
Hence, for any frontier, there are at most $Cp$ operations that can
extend the frontier. Hence, the number of outcoming edges from each
frontier is bounded from above by $Cp$. Recall that the number of
feasible frontiers is at most $O(nC^{p-1})$, the total number of
edges in the frontier graph is $O(nC^pp)$ in the context of pending
periods and physical time orders. Similar to our discussions in the
last paragraph, the above asymptotic order is in fact $O(n)$ for a
given system.

To sum up, with pending period information, frontier analysis
enables one to deal with the pruned frontier graph, which has only
linear numbers of nodes and edges with respect to the number of
operations. Many problems relating to the frontier graph can thus be
simplified. A crucial characteristic of frontier analysis is that it
has successfully utilized the physical time orders, which may
include some orders other than logical time orders, to localize the
computation of many undiscovered orders.

\subsection{Order Analysis: A Technique Beyond Physical Time Order}\label{sec:Order Analysis}
As a new dimension of ordering relations in multiprocessor systems,
the physical time order can order some operations that are
concurrent from traditional points of view. However, it is still
problematic to say that two operations with neither physical nor
logical time orders are concurrent. In this section, we present a
new order that is beyond physical and logical time orders, and then
study the related technique, order analysis, on the basis of the new
order.

As we have mentioned in Section \ref{sec:Physical Time Order}, the
physical time order is independent and consistent with traditional
logical time orders. Hence, it is a natural idea to consider the
combination of physical and logical time orders. Formally, the
transition closure of the physical and logical time orders, which is
now named \emph{time global order}, is defined as follows:
\begin{definition}[(Time Global Order)] We say that operation
$u_1$ is before operation $u_2$ in time global order iff $u_1$ is
before $u_2$ in processor order, or $u_1$ is before $u_2$ in
execution order, or $u_1$ is before $u_2$ in physical time order, or
$u_1$ is before some operation $u$ in time global order and $u$ is
before $u_2$ in time global order. Formally,
\begin{eqnarray}
\nonumber &&(u_1\xrightarrow[]{TGO}u_2) \rightarrow
\big((u_1\xrightarrow[]{PO}u_2)
    \vee (u_1\xrightarrow[]{E}u_2) \vee (u_1\xrightarrow[]{T}u_2) \\
\nonumber
&&\quad\quad\quad\quad\quad\quad\quad\quad\quad\quad\quad\quad\quad\vee(\exists
u\in O: u_1\xrightarrow[]{TGO}u\xrightarrow[]{TGO}u_2) \big).
\end{eqnarray}
\end{definition}
Time global order is not the simple addition of physical and logical
time orders. Due to the transitivity of partial order, two
operations with neither physical time order nor logical time order
may have certain time global order.

Now let us come to a novel technique named order analysis, given the
definition of time global order. Order analysis exploits time global
orders between operations, and utilizes them to check the
correctness of an execution. As we know, the correctness of an
execution in a multiprocessor system is equivalent to whether there
is some cycle in the corresponding execution graph, where the
execution graph is a DAG with its nodes representing operations and
its directed edges representing the orders between operations
(traditionally, these orders are processor order and execution
order). Given the pending period information, a correct execution
must further comply with the time global order. Hence, checking the
correctness of an execution is then equivalent to finding a cycle in
the corresponding execution graph, which contains edges in responses
to processor order, execution order, and physical time order. We
call this type of graph \emph{TGO execution graph} \cite{Chen2009}.

Let $\mathscr{C}$ be the set of all cycles including operation $u$
in the TGO execution graph. Furthermore, let $\mathcal{C}$ be a
cycle belonging to $\mathscr{C}$ ($\mathcal{C}\in\mathscr{C}$), such
that $u$ is an operation in $\mathcal{C}$ ($u\in\mathcal{C}$).
Intuitively, for any operation $u$, there are three kinds of cycles
containing $u$: 1) all operations of the cycle except $u$ are not in
the pending period of $u$; 2) some operations of the cycle are not
in the pending period of $u$, while other operations are in the
pending period of $u$; and 3) all operations of the cycle are in the
pending period of $u$.

Concerning the first kind of cycles which mainly involves operations
outside the pending period of $u$, the following lemma proves that
they can be reduced to specific local cycles involving two
operations.

\begin{lemma}{\cite{Chen2009}} Given a time global order cycle $\mathcal{C}$
containing operation $u$, if all operations in $\mathcal{C}$ except
$u$ are before $u$ in physical time order, there must be a write
operation $w$ in cycle $\mathcal {C}$, which is after $u$ in
execution order. Formally,
\begin{eqnarray}
\nonumber&&\big(\forall v \in \mathcal{C}: (v \neq u) \rightarrow
(v\xrightarrow[]{T}u)\big)\rightarrow (\exists w \in \mathcal{C}:
u\xrightarrow[]{E}w).
\end{eqnarray}
\end{lemma}
\textbf{\emph{Proof.}} Given that operation $v'$ is the successor of
$u$ in cycle $\mathcal {C}$, there are three situations for us to
consider: $u\xrightarrow[]{T}v'$, $u\xrightarrow[]{PO}v'$, and
$u\xrightarrow[]{E}v'$. We know that all operations in $\mathcal
{C}$ except $u$ are before $u$ in physical time order, therefore
$u\xrightarrow[]{T}v'$ does not hold. Since $t_p(v')$ is before
$t_p(u)$, $u\xrightarrow[]{PO}v'$ does not hold. Hence,
$u\xrightarrow[]{E}v'$ holds. Furthermore, if $v'$ is a read
operation, $v'$ certainly cannot get the value of $u$ from the
future, and $u$ cannot be before $v'$ in execution order. Therefore
$v'$ is a write operation. Thus the theorem is proved. $\square$

\begin{figure}[htbp!]
\begin{center}
\epsfig{file=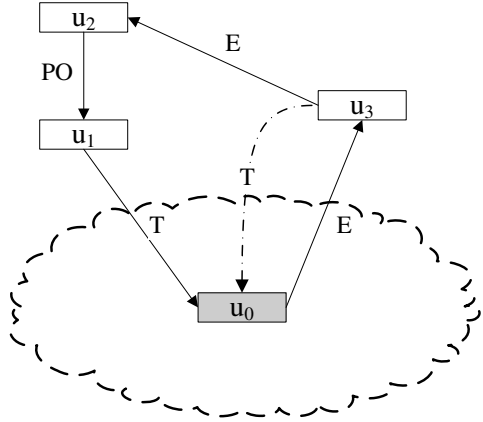, width=0.5\textwidth}
\end{center}
\caption{The operations $u,u_1,u_2$ and $u_3$ consist a cycle in TGO
graph, where $u_1,u_2$ and $u_3$ are before the pending period of
$u$ (the clouds in the figure).\label{figure:first cycle}}
\end{figure}

As shown in Figure \ref{figure:first cycle}, the operations
$u,u_1,u_2$ and $u_3$ consist a cycle in TGO graph, where $u_1,u_2$
and $u_3$ are before the pending period of $u$. Based on the
previous lemma, the cycle can be reduced to a small cycle
$u\xrightarrow[]{E}u_3\xrightarrow[]{T}u$.

Similarly, the second kind of cycles can be reduced to cycles
involving merely an operation outside the pending period of $u$.
Moreover, the third kind of cycles are obviously local cycle inside
the pending period of $u$. Based on the above locality of cycles in
TGO execution graph, we propose three correctness rules in Theorem
\ref{theorem:checking rule} to check the acyclicity of a TGO
execution graph locally. Each correctness rule is related to one
kind of cycle mentioned above.

\begin{theorem}[Checking Rules Theorem \cite{Chen2009}]\label{theorem:checking rule}
There is no cycle in the TGO execution graph of the execution iff
for any operation $u$ of the execution, the following three
correctness rules hold:
\begin{eqnarray}
\nonumber &&\textbf{Rule1: }\forall w \in O:
(w\xrightarrow[]{T}u)\rightarrow
\neg(u\xrightarrow[]{E}w);\\
\nonumber &&\textbf{Rule2: }\forall v,v' \in O:
(v\xrightarrow[]{T}u)\wedge
(v'\xrightarrow[]{GO}v)\rightarrow \neg(u\xrightarrow[]{GO}v');\\
\nonumber &&\textbf{Rule3: }\neg\Big(\exists \mathcal{C} \in
\mathscr{C}: \big(\forall v \in \mathcal{C} :
\neg(u\xrightarrow[]{T}v \vee v\xrightarrow[]{T}u)\big)\Big).
\end{eqnarray}
\end{theorem}

\textbf{\emph{Proof.}} ``$\rightarrow$''. We assume that there is no
cycle in the TGO execution graph of the execution. For Rule 1, given
that a write operation $w$ satisfies $w\xrightarrow[]{T}u$,
$u\xrightarrow[]{E}w$ does not hold. Otherwise there will be a cycle
$w\xrightarrow[]{T}u\xrightarrow[]{E}w$. For Rule 2, if operations
$u$, $v$ and $v'$ satisfy $(v\xrightarrow[]{T}u)\wedge
(v'\xrightarrow[]{GO}v)$, then $u\xrightarrow[]{GO}v'$ does not
hold, otherwise there will be a cycle
$v'\xrightarrow[]{GO}v\xrightarrow[]{T}u\xrightarrow[]{GO}v'$. Rule
3 is trivial: since there is no cycle in the whole graph, there
certainly will be no cycle in the pending period of $u$. Hence
``$\rightarrow$'' is proved.

``$\leftarrow$''. We use reduction to absurdity to prove it. Let us
assume that Rules 1, 2 and 3 all hold, but there is a cycle
$\mathcal {C}$. Let operation $u$ be the last performed operation in
the cycle. According to Rule 3, there must be some operation outside
the pending period of $u$. We can travel $\mathcal{C}$ from $u$. Let
$v$ be the first operation before $u$ in physical time order in
traveling $\mathcal{C}$. Since $u$ is the last committed operation
in the cycle, $u\xrightarrow[]{T}v$ cannot hold. Instead, we have
$v\xrightarrow[]{T}u$. If all operations except $u$ in cycle
$\mathcal {C}$ are before $u$ in physical time order, according to
Lemma 1, there must be some operation $w$ such that
$u\xrightarrow[]{E}w$, which contradicts Rule 1. Hence, there must
be some operation in the pending period of $u$. Let $v'$ be the
precedent operation of $v$ in $\mathcal{C}$. As shown in Figure 2,
$u\xrightarrow[]{TGO}v'$ and $v'\xrightarrow[]{TGO}v$. Let the edge
$a\xrightarrow[]{T}b$ be the first physical time order edge on the
path from $u$ to $v$ in $\mathcal{C}$. According to the definition
of physical time order, we obtain that
$t_p(u)<t_p(a)<t_p(b)<t_p(b)$. However, since $u$ is the operation
committed last in the cycle $\mathcal{C}$, $t_p(u)$ cannot be before
$t_p(b)$, and we reach a contradiction, there is no physical time
order edge from $u$ to $v$ in $\mathcal{C}$. As a result,
$u\xrightarrow[]{GO}v'$ and $v'\xrightarrow[]{GO}v$ hold. But
$u\xrightarrow[]{GO}v'\xrightarrow[]{GO}v\xrightarrow[]{T}u$
contradicts Rule 2. Thus ``$\leftarrow$'' is proved. $\square$

In a real system, Rule 1 checks the incorrect performed time: the
performed time of an operation is out of its pending period, thus
its effect can not be observed by some operations with later and
disjoint pending periods. To check Rule 1, one need to check whether
the latest write before $u$ in physical time order has propagated to
$u$. Rule 2 focuses on ordering bugs between operations inside and
outside of the pending period. To check Rule 2, one need to check
all operations before $u$ in global order to find cycles as shown in
Figure 2. Rule 3 focuses on cycles inside the pending period.

Theorem \ref{theorem:checking rule} not only shows how to check
whether a TGO execution graph is acyclic, but also limits the
complexity of checking. That is because checking one operation only
involves at most $Cp$ operations: Rule 1 involves only a constant
number of operations, while checking Rules 2 and 3 we need to travel
through all operations (the number is at most $Cp$, see also Section
\ref{sec:Frontier Analysis}) in the pending period of $u$. As a
consequence, checking the correctness of an execution is with the
complexity of $O(nCp)$.

Order analysis is beyond both physical and logical time orders,
since it has employed the time global order. This approach is
effective for tackling not only correct behaviors but also error
behaviors of multiprocessor systems (e.g., cycles in execution
graph).

\section{Applications}
In this section, we present two examples, memory consistency
verification and event ordering problems, to illustrate the
significance of our proposed concepts and approaches.

\subsection{Memory Consistency Verification Problem}
In most modern multiprocessor systems, the memory subsystem employs
complex hierarchies with many hardware resources to support shared
memory, which may contain bugs about memory consistency and cache
coherence. Furthermore, parallel programs and compilers may also
bring violations to memory consistency. Even subtle bug about memory
consistency may lead to erratic behavior, which is difficult to find
and debug. The memory consistency verification problem aims at
checking the execution of parallel program against given memory
consistency, and has been widely concerned in both academic and
industrial fields. Gibbons and Korach proved that the VSC problem
(verifying sequential consistency) is NP-hard with respect to the
number of memory operations \cite{Gibbons1992}. Further, they also
studied the complexity of the VSC problem with some additional
information and constraints \cite{Gibbons1994,Gibbons1997}, and
their results include: 1)With read mapping which maps every read to
the write sourcing its value, the obtained VSC-read problem is still
NP-complete. 2)With total write order which orders the write
operations for each memory location totally, the obtained VSC-write
problem is also NP-complete. 3)With both read mapping and total
write order, the obtained VSC-conflict problem belongs to P.
However, read mapping imposes restrictions to the executing parallel
program, while obtaining the total write order even requires adding
specialized hardware \cite{Meixner2005,DeOrio2009}. Some
investigations manage to obtain a relatively low complexity at the
cost of losing completeness (these methods still require read
mapping information)
\cite{Hangal2004,Manovit2005,Manovit2006,Roy2006}. Nevertheless,
without total write order information, even the method with least
time complexity among these incomplete methods still require the
time complexity $O(n^3)$ \cite{Manovit2006}. Henzinger \emph{et al.}
\cite{Henzinger1999} sought natural restriction on multiprocessor
systems, which bounds the number of pending operations in the
system, to enable formal verification of a high-level system
description against memory consistency. However, the method aims at
small and manually-constructed system models and is incapable to
solve real design.

\begin{algorithm}[htbp]
\caption{{\small Algorithm Framework of Memory Consistency
Verification}} \label{table:memory consistency}
\begin{tabular}{lc}
\hline
\quad\quad\quad\quad\quad\quad\quad\quad\quad\quad\quad\quad\quad\quad\quad\quad\quad\quad
\quad\quad\quad\quad\quad\quad\quad\quad\quad\quad\quad
\end{tabular}
\algname{\large{int memory\_consistency\_verification}}{}
 \begin{algtab}
     $f=starting\_frontier$;\\
     \algwhile{$1$ \textbf{do}}
         \algif{\algcall{checking\_violence}{$current\_execution\_graph$} \textbf{then}}
               \algif{$f==starting\_frontier$ \textbf{then}}
                         \algreturn{0};\\
               \algend{\textbf{end if}}\\
               \textbf{else}\\\algbegin
                         \algcall{remove\_edge}{$current\_execution\_graph$};\\
                         $f=\algcall{backtrack}{f}$;\\
               \algend{\textbf{end else}}\\
         \algend{\textbf{end if}}\\
         \textbf{else}\\\algbegin
               \algif{$f==terminating\_frontier$ \textbf{then}}
                         \algreturn{1};\\
               \algend{\textbf{end if}}\\
               \textbf{else}\\\algbegin
                         $f=\algcall{select\_branch}{f}$;\\
                         \algcall{add\_edge}{$current\_execution\_graph$};\\
               \algend{\textbf{end else}}\\
         \algend{\textbf{end else}}\\
    \algend{\textbf{end while}}\\
    \end{algtab}
    \begin{tabular}{lc}
\hline
\quad\quad\quad\quad\quad\quad\quad\quad\quad\quad\quad\quad\quad\quad\quad\quad\quad\quad
\quad\quad\quad\quad\quad\quad\quad\quad\quad\quad\quad
\end{tabular}

\end{algorithm}
In our previous work \cite{Chen2009}, we proposed a fast and
complete memory consistency verification method. Our method requires
neither read mapping nor total write order, which makes our method
easy to generalize. This method only needs to observe the pending
periods of part of operations periodically to assign a pending
period to each operation using the assignment analysis approach
presented in Section \ref{sec:Assignment Analysis}. As shown in
Table \ref{table:memory consistency}, the framework of the algorithm
in our method is inherited from \cite{Gibbons1994}, which requires
to traverse the frontier graph to find a path from the starting
frontier to the terminating frontier. At each frontier (node of the
frontier graph), the checking\_violence function checks cycle in the
current execution graph: If any cycle is found in the current
execution graph, the algorithm should backtrack to the previous
frontier and select another branch to explore; if the current
execution graph is proven to be acyclic, the algorithm should move
forward to a next frontier. Each time moving forward in the frontier
graph, some edges of execution order should be added to the current
execution graph; each time backtracking in the frontier graph, some
edges of execution order should be removed from the current
execution graph. Once the current frontier $f$ travels to the
terminating frontier and no cycle is found, the execution is proven
to comply with memory consistency model. If the current frontier $f$
has backtracked to the starting frontier and no other branch can be
selected, the execution is proven to violate the memory consistency
model.

The complexity of memory consistency verification comes from the
product of two aspects. One aspect is the complexity of traversing
the frontier graph, and the other aspect is the complexity of
checking violence in the current execution graph at each frontier.
Recall that with pending period information, the frontier analysis
presented in Section \ref{sec:Frontier Analysis} bounds the numbers
of frontiers in the frontier graph to $O(nC^p)$ from above, and the
order analysis bounds the complexity of checking cycle in execution
graph to $O(nCp)$, therefore the overall time complexity for
complete memory consistency verification is only $O(n^2C^pp)$.

It is worth noting that in error multiprocessor systems, there may
be bugs of improper performed times for operations: the actual
performed time of an operation may not been globally observed before
the obtained end time of the operation. For example, in a directory
based cache coherent system, a store may not be observed by other
processors because of some error in the directory, thus its actual
performed time is later than the obtained end time. However,
although our memory consistency verification method expects that the
obtained pending period of each operation should contain its actual
performed time, it does not require the certified precondition that
the obtained pending period of each operation contains its actual
performed time (though we have defined ), since it can find
violations of both memory consistency model and improper performed
time.

On the basis of our theoretical investigations, we have implemented
a memory consistency verification tool for CMP, which is named
LCHECK \cite{Chen2009}. LCHECK can verify a number of memory
consistency models \cite{Steinke2004}, including sequential
consistency, processor consistency, weak consistency and release
consistency. It has become an important verification method for the
functional validation of an industrial CMP called Godson-3
\cite{Hu2008,Hu2009}, and have found many bugs of the memory
subsystem of Godson-3.

\subsection{Event Ordering Problems}
Event ordering is another interesting topic related to
multiprocessor systems. In different candidate executions of one
parallel program, two operations in the program may occur in
different orders. The event ordering problems investigate the
inevitability or possibility of the order between two operations.
They are the theoretical foundations of many other important
problems in multiprocessor system, such as replay of execution
\cite{Hower2008}, debugging software \cite{Pozniansky2003,Lu2006},
intrusion detection \cite{Leblanc1987}, and so on.

In \cite{Netzer1990}, Netzer and Miller gave a formal analysis for
the event ordering problems. They defined the happened-before,
concurrent-with, and ordered-with relations, for operations. Each of
the three ordering relations was further defined in two manners:
must-have and could-have (similar with universal quantifier and
existential quantifier in symbolic logic). The must-have sense
requires that the ordering is guaranteed in all legal candidate
executions of the program (with respect to the given memory
consistency model), and the could-have sense requires that the
ordering occurs in at least one legal candidate execution of the
program. Netzer and Miller found that it is co-NP-hard to prove any
of the must-have ordering relations and it is NP-hard to prove any
of the could-have ordering relations. For the sake of brevity, here
we only discuss the must-have happened-before (MHB) and could-have
happened-before (CHB) relations in detail, while the discussions of
other relations are similar to the two examples.

According to Gibbons and Korach \cite{Gibbons1994}, one path from
the starting frontier to the terminating frontier on the frontier
graph can represent a candidate execution of the program, the event
ordering problems, which depend on the candidate executions, can be
investigated on the frontier graph of the parallel program. Assume
that operation $u$ is an operation of the processor $\mathcal{P}_i$,
operation $v$ is an operation of the processor $\mathcal{P}_j$, and
$\mathbb{P}$ is the set of all paths from the starting frontier to
the terminating frontier on the frontier graph, then the must-have
happened-before (MHB, denoted by ``$\xrightarrow[]{MHB}$'') and
could-have happened-before (CHB, denoted by
``$\xrightarrow[]{CHB}$'') relations can be formalized as follows:
\begin{eqnarray}
u\xrightarrow[]{MHB}v\leftrightarrow (\forall P\in \mathbb{P}:
(\exists f\in P: (\mathcal{P}_i(f)\xrightarrow[]{PO}u)\wedge
(v\xrightarrow[]{PO}\mathcal{P}_j(f))));\\
u\xrightarrow[]{CHB}v\leftrightarrow (\exists P\in \mathbb{P}:
(\exists f\in P: (\mathcal{P}_i(f)\xrightarrow[]{PO}u)\wedge
(v\xrightarrow[]{PO}\mathcal{P}_j(f)))),
\end{eqnarray}
where $\mathcal{P}_i(f)$ is the operation of frontier $f$ on
processor $\mathcal{P}_i$. Regarding the frontier graph, the
must-have happened-before relation between $u$ and $v$ is that for
each path from the starting frontier to the terminating frontier,
there is a frontier $f$ on the path whose operation on processor
$\mathcal{P}_i$ is after $u$ in processor order and whose operation
on processor $\mathcal{P}_j$ is before $v$ in processor order. Thus
deciding the must-have happened-before relation between $u$ and $v$
is equivalent to a basic problem of the graph theory: to decide all
paths (paths on frontier graph) from one node (the starting
frontier) to another node (the terminating frontier) in a DAG (the
frontier graph) must pass through a set of nodes (the set of
frontiers which imply $u$ happens before $v$), which has time
complexity of $O(n_f+e_f)$ \cite{Chartrand2004} (where $n_f$ is the
number of nodes in the frontier graph and $e_f$ is the number of
edges in the frontier graph). Therefore, the time complexity of the
must-have happened-before problem is bounded from above by some
linear function with respect to the numbers of nodes and edges in
the frontier graph.

Similarly, the could-have happened-before relation of operation
between $u$ and $v$ $u\xrightarrow[]{CHB}v$ is that there exists a
path from the starting frontier to the terminating frontier on the
frontier graph, which contains a frontier $f$ whose operation on
processor $\mathcal{P}_i$ is after $u$ in processor order and whose
operation on processor $\mathcal{P}_j$ is before $v$ in processor
order. It is equivalent to another basic problem of graph theory: to
decide a path (a path in frontier graph) from one node (the starting
frontier) to another node (the terminating frontier) in a DAG (the
frontier graph) which passes through a set of nodes (the set of
frontiers which imply $u$ happens before $v$), which has time
complexity of $O(n_f+e_f)$ \cite{Chartrand2004}. Hence, the
complexity of the could-have happened-before problem is bounded from
above by some linear function with respect to the numbers of nodes
and edges in the frontier graph.

From the above discussions, we can find that the complexities of the
two event ordering problems directly relate to the numbers of nodes
and edges in the frontier graph. Similar discussions can be
generalized to other event ordering problems. Given the restrictions
of pending periods and physical time orders, the complexities of
event ordering problems can both be reduced to $O(nC^pp)$, since
both the nodes and the edges of frontier graph is no more than
$O(nC^pp)$ with the additional information.

\section{Conclusion}
In this paper, a novel perspective of utilizing global clock in
multiprocessor systems is presented, demonstrating that the
implication of global clock, if being exploited sufficiently, can
have significant influence on the design and analysis of
multiprocessor systems. As we have pointed out in Section
\ref{sec:introduction}, a global clock in a multiprocessor system
actually implies a physical-time-based partial ordering for all
operations in the system. It is revealed by our investigations that
the above partial ordering can not only export useful information
for supplying logical time order information, but also provide
natural constraints for localizing the inference between operations.
Such natural constraints are defined explicitly as a partial order
named physical time order, which has been proven to be independent
and consistent with traditional logical time orders.

On the basis of the above views and concepts, we have proposed a
number of approaches, which are named pending period analysis as a
whole, focusing on different aspects of making our idea of utilizing
global clock practical. These approaches, together with the
definitions of pending period and physical time order, actually
provide solutions for the difficulties mentioned at the end of
Section \ref{sec:motivation}. Concretely, the concept of pending
period, which is actually a flexible relaxation of the precise
performed time, has given a feasible solution for handling the
hard-to-obtain precise performed time. Moreover, the frontier
analysis presented in Section \ref{sec:Frontier Analysis} has
limited the complexity of conjecturing or inferring the ordering
relations through pruning the space of candidate executions, and the
order analysis presented in Section \ref{sec:Order Analysis} further
combines the physical and logical time orders to infer the ordering
relations inside overlapped pending periods. Finally, the assignment
analysis carried out in Section \ref{sec:Assignment Analysis}
demonstrates that, observing pending periods of only part of the
operations is enough to obtain pending periods of all operations.

The pending period analysis has been adopted in two application
problems in multiprocessor systems. One of these problems, complete
memory consistency verification, is simplified from NP-hard to the
time complexity of $O(n^2C^pp)$ with pending period analysis. This
fast and complete memory consistency verification method has been
employed in industry. Moreover, the two event ordering problems,
which were proven to be Co-NP-Hard and NP-hard respectively, can now
be solved with the time complexity of $O(nC^pp)$ if restricted by
pending period information. It can be hoped that more problems in
multiprocessor systems can be facilitated by the view, concepts and
approaches proposed in this paper.

%\section*{Acknowledgment}
\bibliographystyle{}

\end{document}